# Information Transfer in the Agricultural Sector in Spain

Abstract:

This paper examines the structures of information transfer to the agricultural (production) and agro-alimentary (transformation and commercialization of the products) sector within Spain. A historical perspective is provided to better illustrate the reality and complexity of Spain with regard to the systems of agrarian extension, agricultural research, resources provided by Spain's central administration, and the use of information by related enterprises.

The Service of Agrarian Extension appeared in Spain in the 1950s, and new political-administrative structures (agribusiness associations, cooperatives) were founded when Spain became a democratic nation in the late 1970s, and with the arrival of electronic information, largely in the 1990s. We also describe the tools supporting innovation in the agro-alimentary sector: centers of agrarian research and technological centers. Finally, reference is made to the means of communication dedicated to the agricultural sector.

The paper illustrates that the systems of agricultural information in Spain have been largely derived from initiatives of the Public Administration, with few private initiatives.

# INTRODUCTION

A study of information systems within the Spanish agricultural sector should begin with a brief overview of the agrarian situation in Spain over the past century to provide a better context and perspective for understanding the evolution occurring in the country as a whole.

Until recent decades, the economic history of Spain has been, to some extent, the history of the agrarian sector. It was not until the 1950s when the Franco government designed the National Stabilization Plans as a mechanism for modernizing the economy, and, therefore, the country as a whole. Spain was essentially an agrarian country until the fifties, with the exception of limited industrial development in peripheral zones (Cataluña and the Basque Country).

The agrarian sector, based on low-yield crops, mostly for domestic consumption, maintained these characteristics until the '50s. The process of modernizing this sector was interrupted by the Spanish Civil War (1936-1939) and by the first period of the government of Franco, which came to be known as the *Autarquía*, in which Spain shut its borders to foreign investment and to the entry of merchandise, services and capital. In this period of history, the technical and social conditions of rural Spain were not typical of the $20^{th}$ century, with very scarce technical resources and abundant unqualified labor. Then, after the fifties, a timid transformation began in the countryside which also furthered industrial development: agriculture provided the work force (an exodus of farmhands from the country to the city), the capital (thanks to the savings of farmers) and also the goods.

Meanwhile, change was taking place in the agricultural model, with the concentration of plots of land (the fusion of farmlands, but not of property) to make this model more competitive. New techniques and mechanization promised greater productivity. Also around this time, a plan for irrigation was established as a necessary part of modernizing the agrarian sector.

The rise of democracy, with the Spanish Constitution in 1978, spawned a regimen of freedom along with a radical change in the political management of agricultural interests, with the transition from a centralized or federal state to a country made up of autonomous communities; that is, the regions gained political power. So it was that most agricultural policy came under the umbrella of local governmental bodies. Hence, after 1979, there was a rapid shift of power, from the State administration toward the regions[1]. Authority regarding agriculture and livestock was transferred (obligated by article 148.1.7º of the Constitution) along with responsibilities in areas as diverse as the recovery and extension of farmlands, crop sanitation, animal sanitation, agricultural research, nature and conservation, farm chambers, agricultural reform and development, wine-making, agrarian commercialization, the agrarian societies of transformation, farm industries, and agro-alimentary quality. The regions managed subventions from the *Fondo Español de Garantía Agraria (FEGA)*[2], although the state remained in charge of the management of funds through the FEGA for national responsibilities such as the management and payment of restitutions for exports, food to the needy, nutritional promotion, etc.

Meanwhile, Spain joined the European Economic Community (1986) —the forerunner of the European Union— with its very well defined agricultural policies through the Common Agrarian Policy known as CAP, which led to the closure of less profitable operations, yet maintained the degree of protectionism with respect to other countries. This made Spain's agrarian sector, within the European framework, highly dependent upon government funding. All these changes caused variations in the transfer of information, both in form and in substance.

Agricultural activities (including livestock and fishing) accounted for 2.6% of the bulk economy of Spain (GNP) in 2007 (INE, 2008, p. 29). This very lean figure reflects the paradigmatic change of Spain's economy over the past 70 years. The most recent *Survey on the Structure of Agrarian Holdings* conducted by the National Institute of Statistics (INE, 2008b) to analyse the state of Spanish agriculture and document the structural evolution of agricultural operations (data refer to the farm campaign of 1 October 2006 to 30 September 2007) confirmed a trend toward decreasing numbers of farms and an increase in the agricultural surface utilized. The number of farms dropped 3.3% with respect to the previous year, while the surface utilized increased by 0.2%.

---

[1] The process of decentralizing Spain politically into autonomous communities was by no means a simple task, and many problems arose. The transferences led to a situation in which legal regulation in each one of the 17 regions was different in many cases, generating problems within one state with specific norms for each territory. Some parties responsible for agricultural business organizations have requested that certain competences be given back to the central government, such as water and animal health. They explain that, for example, a livestock owner who wants to sell his calves to buyers from other regions must fulfill different prerequisites, depending on the region they are sold to.

[2] The Spanish Fund for Agrarian Guarantee or *Fondo Español de Garantía Agraria* (FEGA) is an autonomous organism, ascribed to the Ministry of Agriculture, which processes agricultural and livestock policy, and whose main function is to manage the aid that comes through the Common Agricultural Policy (CAP). In Spain, the competences of the region make each one become a managing and paying agent for this aid, even though the State, through the FEGA, carries out the activities of coordination and centralization of the information before the European Union. The CAP funds are financed by means of the FEAGA (Fondo Europeo Agrícola de Garantía Agraria) and FEADER (Fondo Europeo Agrícola de Desarrollo Rural).

The GNP of industry and energy —which would include agro-alimentary businesses and the transformation of products— was 15.7% in 2007. Of the total GNP of Spain's industry, the agro-alimentary sector is the one with the greatest volume, 15.4% of the total Data from 2009 show over 30,000 firms active in the food and drink industry, although more than 16,000 firms had no steady employees or only one or two and some 8,000 firms had from three to nine workers.

## AGRICULTURAL INFORMATION: FROM EXTENSION TO THE INTERNET

**Antecedents of the Service of Agrarian Extension**

Spanish researcher Sánchez de la Puerta (1996) tells us that, until 1955, there was no agreement in Spain as to how Extension[3] might be carried out in an organized and structured manner from the State powers, unlike in other European countries and the United States, where the extensionist practice —which gave rise to agencies making up the national services of Extension— can be traced back to the end of the 19th century and depending on the historical period in question, the responsibility for Extension in Spain was held by teachers, parishioners, and farming engineers who made use of Agricultural Treaties, farm-schools, the *Cartilla* (a brief agricultural treaty) and the Catechism of Agriculture (frequent questions and answers about the agrarian sector), as obligatory reading in schools. Besides Sunday lessons by teachers and priests, there were publications; exhibitions; experiments such as the *Servicio de Aplicaciones Agrarias*, created in Soria in 1928; and the activities of the *Servicio de Divulgación y Propaganda Agrícola* during the 1940s. All of these instruments —and especially the farm-schools— set precedence for the Service of Agrarian Extension as it took root in Spain.

---

[3] The term *Extensión* comes from the service instituted in the United States, the Cooperative Extension Service, borrowing the term extension from the English tradition, where it was used to define the activities carried out by a University beyond its campus (Jones, 1981). In Spain, this term has been substituted by others, such as divulgation or propaganda, used in the denomination of different public organisms. Sánchez de la Puerta points out that in the 18th century the then minister Campomanes used the term Extensión for the transmission of new agricultural knowledge.

Furthermore, the Extension practices, according to some authors, responded to different strategic needs. Thus, "the socio-political currents previous to the Spanish Civil War of 1936-1939 viewed Extension as an instrument for the technical progress of agriculture. For the Catholic Agrarian Syndication and Falange, capitation and agrarian divulgation were at the same time instruments for the control of the rural population and for the increase of farm productivity; whereas for the national Movement, Extension was useful for maintaining farmers in their realm and was a means of preserving the values where they deposited their ideology" (Sánchez de la Puerta, 1996, p.389). Under Franco, Extension was used as a tool for economic development that also perpetuated a certain ideology. It was common to find Extension-related activities carried out by the Youth Front or by the Feminine Section (*Frente de Juventudes* and *Sección Femenina*, both organisms ascribing to the Franco regime) in which young men and women were respectively active.

Until 1955 however, all these activities of Extension in Spain were dispersed. The activities generated by the farm-schools, the Catholic Syndication, or those undertaken by the different organisms under the Ministry of Agriculture were carried out in a largely unconnected manner. Gómez Torán (1982) explains that this lack of connection led to a perceived deficiency in government efforts to provide information to improve agriculture and the quality of life of the rural population.

**Creation of the *Servicio de Extensión Agraria* (SEA)**

Thus, the Service of Agrarian Extension can be seen as the Spanish version of the North-American Cooperative Extension Service. The Service was brought to Spain by the hands of North-American experts in Extension, in a climate of cooperation and aid. As we know, the U.S. initiated a plan for aid to Europe known as the Marshall Plan, which did not come directly, but rather arrived in Spain via other direct aid programs, under the McCarran Amendment and Public Law 480[4]. The U.S. Economic Mission led to an invitation to Spain's agricultural minister, Rafael Cabestany, to visit the Extension Service of the United States. This trip took shape after the publication of the Ministerial Order of 15 September 1955, leading to the experimental creation of the Spanish Service of Agrarian

---
[4] Before, in 1953, the Pact of Madrid was signed. This meant the partial international recognition of the Regime of General Francisco Franco. The Pact contained bilateral agreements in defense, economic cooperation, and technical assistance, and allowed for the establishment of U.S. military bases in Spain.

Extension and a trip to the United Status by an entourage of Spanish technicians in charge of learning details about how the U.S. Service worked. The Spanish SEA reached beyond the scope of the *Instituto Nacional de Investigaciones Agronómicas* (a research organization accused in its early years of producing no real research), so that the innovations diffused by the SEA could, in fact, be considered the results of U.S. efforts.

The system was organized around regional agencies (each region consisting of several municipalities), with two or three agents and absolute autonomy. The agents held undergraduate technical degrees and acted as monitors. Aside from their broad technical preparation, they were educated in rural sociology, group dynamics, agrarian association, and techniques of communication and publication (Martínez Costa y Claudio Franco, 2005, pp. 73-75).

The ultimate objective was to improve the productivity of agrarian operations in order to improve the lifestyle of rural populations. As political objectives, the rural aid programs aspired to the educational, social and economic enhancement of these individuals. The SEA, through its agencies, maintained a constant flow of information through the following means: responding to consultation within the agencies; visiting operations in the countryside; demonstrating methods; holding meetings and informative talks; and organizing groups, discussions, or trips to farms. In the 1970s, its work was structured into four blocks:

– Work carried out in the farms themselves, through the diffusion of technology and managerial formation, where involvement of SEGEs or Seminars of Extension was usual.
– Work with youth, done through Extension academies where the offspring of farmers and the *Planteles* of Extension (teams of young adults) received general training.
– Work with the household.
– Cooperative efforts and community development (Sánchez de la Puerta, 1996, p. 423).

The Beginning of the End of the Service of Agrarian Extension

After the 1970s, the SEA model entered a critical period due to the administrative break-up of Spain into Autonomous Communities, so that means and personnel were moved to the regions. Even before this however, since the beginning of the seventies, tension ran high between those agents more concerned with the traditional work of Extension and the diffusion of innovation or economic development of the rural sector and those for whom social development prevailed over technical aspects. The transfer to the *Comunidades Autónomas* of all the Extension responsibilities that had previously come under State power also caused a change in the internal model, as one of the first consequences was that the agrarian research policies and the Extension policies that had been separate would now be united.

The final phase of the SEA as a State organism coincides with the process of negotiating the entry of Spain into the European Community. At that turning point, the modernization of Spanish agriculture loomed as a crucial necessity and the diffusion of technological information as the most important ingredient. So it was that, in 1981, the National Plan for the Technological Diffusion for the Modernization of Agrarian Structures was launched which stated that "modernization and adaptation of Spain's agriculture calls for deep improvement of agrarian activity, requiring that scientific advances and available technical knowledge be effectively divulged among farmers and that their efficient application to structures be promoted, which is where they acquire their final utility".

With the transference of Extension responsibilities to the regions, persons working at the SEA headquarters became in charge of managing European agricultural funds, at least initially. By virtue of a Royal Decree in 1991 (*Real Decreto* 6454/1991), the SEA disappeared as an autonomous organism within the Ministry of Agriculture.

The CAP —created in the 1960's when Europe was severely lacking in alimentary goods as a means to support prices and ensure supply and, therefore, uphold productivity and improve the standard of living of rural populations— was faced with profound transformation. The increase in agricultural surpluses and concern about the impact that protectionist European policy was exerting upon the less developed member states gave rise to a sounding reform in 1992. The measures aimed to support prices were generally substituted by direct aid, without overlooking support for rural areas. Two clearly differentiated stages in the wake of Spain's democracy may be delineated:

- First, an informational "revolution" entailing the autonomous community shift brought about by the Spanish Constitution of 1978 and the progressive dismantling of the Service of Agrarian Extension, with a search for more effective diffusion of technological information.

- A secondary "revolution" deriving from Spain's membership in the European Economic Community, in which information about aid for the agricultural sector is predominant. This second stage can be characterized by the definitive assumption of Extension responsibilities by regional entities, though also meriting mention is the decisive arrival of new actors that will carry out important Extension work, such as the professional agrarian associations or the farm cooperatives.

## THE DEVELOPMENT OF ELECTRONIC INFORMATION

The development of the agricultural and agro-alimentary sector, like all other economic sectors, has benefited from the electronic information industry. Many recent studies document the value of databases to agri-business research (Salisbury y Tekawade, 2006) and/or promote specific types of resources, such as market studies, as essential for a knowledge of agri-business markets and the development of new products (Smith, 2007). A review of electronic information for the agricultural and agro-alimentary sector in Spain is may be divided into four basic sub-sections: (a) installation of the first computerized information systems, (b) the National Statistic Plan, (c) geographical information systems, and (d) other systems.

**The Early Systems**

In the 1980s we find the first databases for the agricultural sector, which use highly diverse channels of information distribution, in a stage prior to the expansion of the Internet. Information is distributed by means of the Basic Telephone Network (*Red Telefónica Básica,* or RTB); the Iberpac network (a Spanish network meant to transmit information by telematic means); and by videotex, known in Spain as Ibertex, a technology that began to surge forward in the 1990s, but quickly succumbed to the superiority of the Internet.

The Ministry of Agriculture, Fishing and Foodstuff was to design a system called "Agri-pyme" made up of various component databases: agro-meteorology (with daily meteorological predictions for the whole country); agrarian productions and advances (following the forecasts for agricultural products); insect pest warnings (and information about their treatment with pesticides); compensatory mechanisms, prices and markets (for agricultural products, livestock, fishing and forestry products); and animal health (sanitary situation, vaccines and requirements for fish movements). Additional databases were dedicated to support small enterprises, like institutional aid, public bidding and European fairs.

In the decade of the 1980s, some regions like Andalucía —a large autonomous community which had assumed authority in agricultural matters— began to produce databases on their own. The *Instituto de Fomento de Andalucía* created a database on agrarian research projects, while the Environmental Agency of Andalucía (*Agencia del Medio Ambiente de Andalucía*) produced the Environmental Information System of Andalucía. In the private sector, less developed than the public one, deserving mention is the Telebroker firm, with a database covering a number of farming and fishing-related topics.

The incipient electronic information industry of the 1980s can be studied in Spain through the catalogs published by Fuinca (*Fundación para el Fomento de la Información Automatizada*) and Fundesco (*Fundación para el Desarrollo de la Función Social de las Comunicaciones*). Among these are those of the videotex sector which, as mentioned earlier, underwent a bit of a boom in Spain. Before its disappearance (due to the Internet), some of the informational products were commercialized. The Catalog of Ibertex Services (Spain's videotex) of 1993 (Ruiz González, 1993) contains 15 services offered by videotex, some of them with free access, others entailing a fee. One of the most important services was Agritel (which included the aforementioned Agri-Pyme), featuring a great variety of databases. We also begin to see some diversification in the set of information providers: there are professional agrarian organizations, cooperatives, business associations, financial entities, professional colleges of veterinarians, and private firms that offer a wide array of information, from the prices of agricultural products to information about aid, legislation, or plant and animal sanitation. The Catalog of 1995 (Ruiz González, 1995) offers a total of 25 services, with new incorporations including the National Institute of Agrarian Research and Technology (*Instituto Nacional de Investigación y Tecnología Agraria,* or INIA).

The Internet's arrival in Spain made waves among citizens and Administrations in the mid-nineties. It provoked such convulsion in the informational sector that the scarce industry of electronic information —which adapted rapidly to the Internet as the means of distribution was joined by a huge conglomerate of organizations who foresaw that the Web would prove to be an easy means of reaching audiences who were previously difficult to access.

As affirmed by Glynn and Koenig (1995), the Internet stands as a dramatic change for the information environment of the Pymes, as they find a multitude of new providers joining the electronic information industry. At the same time, novel systems of information retrieval arise, eliminating the search engine systems of each information provider —which had to be learned one by one— and new information brokers appear. In short order, the Governments themselves adapt to the new media as providers of information for businesses, immersed in the new concepts of e-Government and e-Democracy, in which informational transparency becomes the object of desire.

In sum, the nineties bring a much greater visibility of information in electronic format —information that had been previously hidden in the nooks and crannies of Administrations in paper form. In this context, the European Union points to the problem of public information not in terms of volume, but rather in terms of accessibility. Its 1998 Green Paper on Public Sector Information in the Information Society emphasizes that "The issue at stake is not that Member States should produce more information, but that the information which is already available to the public should be clearer and more accessible to potential users" (Comisión Europea, 1998, p. 1).

The information sources that relied on videotex abandon this medium and begin to utilize Internet applications. First they adapt to Gopher and then, en masse, to the World Wide Web.

The presence of Spain's Ministry of Agriculture on the Internet can be traced to the early months of 1998 (Muñoz Cañavate, 2003); from then on, as in other countries, the successive governments would initiate information policies with the Internet as the consensual star medium.

**The National Statistic Plan and Statistical Operations**

- Agrarian Census
- Survey of the Structure of Agricultural Exploitations
- Survey of the Methods of Production in Agricultural Exploitations (in process)
- Survey of Territorial Segments
- Calendars of Sowing and Harvest
- Monthly Forecasts of Surface Area and Agricultural Productions
- Surfaces Area and Annual Productions of Crops
- Winegrowing Surveys
- Livestock Forces (Directories and Surveys)
- Statistics of Livestock Production (Directories and Surveys)
- Monthly and Annual Surveys of Incubation Rooms
- Statistics of Rabbit Breeding
- Monthly Evaluation of Commercial Livestock Movement (MOCOPE)
- Utilization of Means of Production
- Annual Statistics of Other Forestry Uses
- Annual Hunting Statistics
- Annual Fluvial Fishing Statistics
- Statistics of Production and Commercialization of Reproductive Forest Material
- Annual Statistics of Lumbering
- National Forest Inventory
- Annual Statistics of Forestry Projects and Actions
- Sustainable Forestry Management
- Current Prices of Agricultural Products
- Current Prices of Livestock Products
- National Token Prices
- Monthly and Annual Statistics of Agrarian Prices and Wages
- Average Annual Prices of Lands of Agricultural Use
- Annual Rules for Rural Leasing
- Balances of Provisions of Agro-livestock Products
- National Balance of Work
- National Agrarian Accounting Network (RECAN)
- Economic Accounts of Agriculture

|  |
|---|
| - National Economic Accounts of Silviculture<br>- Regional Economic Accounts of Silviculture |

Geographic Information Systems

Spain's Ministry of Agriculture has various geographic information systems, as well as vast databases in which information of a spatial nature appears (Table 2):

- *Sistema de Información Geográfica de Parcelas Agrícolas* (SIGPAC) [Farming Land Geographical Information System], which allows for the geographic identification of land parcels declared by farmers and ranchers, in any aid scheme related to surface areas cultivated or used by livestock.

- *Sistema de Información Geográfico Agrario* (SIGA) [Agrarian Data Geographic Information System], which allows for the visualization of cartographic information (elevation, agrarian and livestock regions, coastlines, fluvial boundaries of hydrographic basins), agro-climatic information (temperature, rainfall), and crop maps.

- *Sistema de Identificación de Instalaciones de Acuicultura*, [Identification System for Fish Farming Facilities], which allows for the geographic localization of the continental and marine fishery establishments of Spain.

- *Sistema de Información Geográfica del Área del Medio Rural y Marino* (SIGMAPA) [Geographic Information System for Rural and Marine Affairs], which contains information about irrigation, crop maps, agro-climatic data, denominations of origin, agro-alimentary industries, processing of aid, hydrographic basins, general cartography, etc.

| | |
|---|---|
| Sistema de Información Geográfica de Parcelas Agrícolas (SIGPAC) | http://sigpac.mapa.es/fega/visor/ |
| Sistema de Información Geográfico Agrario (SIGA) | http://sig.mapa.es/siga |
| Sistema de Información Geográfica del Área del Medio Rural y Marino (SIGMAPA) | http://sig.mapa.es/geoportal/ |
| Sistema de vigilancia y alerta de la evolución anual de los cultivos y aprovechamientos | http://www.mapa.es/es/sig/pags/cultivos_apr/index.htm |
| Sistema de Identificación de Instalaciones de Acuicultura | http://webserv.mapa.es/visoracuicultura/matlab/gmf_apps/butm/utm.phtml |

Other Systems

In 2007, Law 11/2007 regarding access by citizens to public services was passed. This law, better known as the Law of Electronic Administration, positioned Spain among the leading countries in this realm, at least in legislative terms, as the Law made it obligatory (no longer optional) for all public administrations to offer telematic services in lieu of or to complement existing services. Along with this norm came standards of quality control for all State entities (R.D. 951/2005), as well as the creation of the State Agency for the Evaluation of Public Policy and the Quality of Services As a result, the Ministry of Agriculture set up numerous informational resources, concerned with legislation, aid and funding, economics, plant and animal health, and quality standards. In the livestock sector two new information systems came into service between the Central Administration and the regional governments:

1. *Sistema de Identificación y Movimiento de Ganado (Bovino)* (SIMOGAN) [National Livestock (Bovine) Identification and Movement System], a database identifying all bovine operations Spanish soil, as well as slaughter or export.

2. *Sistema Nacional de Identificación y Registro de los Movimientos de los*

*Porcinos* (SIMOPORC) [National System for the Identification and Registration of Swine Movement], which records all existing porcine operations on Spanish soil, as well as the movement of these animals.

Also deserving mention are the *Observatorio del Consumo y la Distribución Alimentaria* [Food Consumption and Distribution Monitoring Center], which was established to analyze the habits of Spanish consumers and shoppers as well as commercial distribution strategies, and the *Observatorio de Precios de los Alimentos* [Food Prices Monitoring Center], established in 2000 to achieve price stability by promoting transparency and rationality in the setting of food prices for the benefit of producers and consumers alike.

**RESEARCH IN THE AGRICULTURAL SECTOR IN SPAIN**

|  |  |
| --- | --- |
| Public R&D System | Universities and public research organisms |
| Organisms of support for R&D&I | Organisms for the Transfer of results of research<br>Foundations<br>Organisms of promotion, evaluation, financing and prospective<br>Technological Centers<br>Agencies to promote innovation<br>Technological parks and major installations |
| Other instruments of support | Chambers of commerce, industry and navigation<br>Business organizations |

Source: FECYT

**Public Research in the Agricultural Sector**

At the statewide level, and in direct relation with the agricultural and agro-alimentary sector, are two major public research organisms: the Upper Council of Scientific Research

(*Consejo Superior de Investigaciones Científicas,* or CSIC, with highly diverse centers) and the National Institute of Agricultural and Alimentary Research and Technology (*Instituto Nacional de Investigación y Tecnología Agraria y Alimentaria*, or INIA). Because authority for agricultural research in the public sector is partly transferred to the regional governments, there are responsibilities that pertain to the state, others that come under the regional or autonomous communities, and still others that should be undertaken in a spirit of cooperation. So as to carry out these collaborative efforts, the Coordinating Commission for Agricultural Research (*Comisión Coordinadora de Investigación Agraria*) was created by virtue of a Ministerial Order on 8 January 1987 under the auspices of the INIA.

**a) The National Institute of Agrarian and Alimentary Research and Technology (INIA)**

The INIA has existed since 1971 and is the result of the fusion of several institutes, some of which go back as far as the beginning of the 20th century. The INIA carries out a dual function with regard to agro-alimentary R&D&I. On the one hand, it performs management tasks (programming, coordination, allotment of resources, follow-up and evaluation) for activities related to scientific and technical research. On the other hand, it executes research functions and technological development, aside from technology transfer in matters related to food and agriculture. In addition, the INIA library, together with the libraries of the regional agricultural research centers, comprise the Network of Agricultural Information and Documentation (*Red de Información y Documentación Agraria*, or RIDA), which includes over 40 libraries dedicated to diverse and specific topics within the areas of agriculture, livestock, forestry, alimentation, etc.

**b) The Regional Research Centers**

| | |
|---|---|
| Andalucía | Instituto Andaluz de Investigación y Formación Agraria, Pesquera, Alimentaria y de la Producción Ecológica (IFAPA) |
| Aragón | Centro de Investigación y Tecnología Agroalimentaria de Aragón (CITA) |
| Asturias | Servicio Regional de Investigación y Desarrollo Agroalimentario (SERIDA) |
| Baleares | |
| Canarias | Instituto Canario de Investigaciones Agrarias (ICIA) |
| Cantabria | Centro de Investigación y Coordinación (CIC) |
| Castilla y León | Servicio de Investigación, Desarrollo y Tecnología Agraria |
| Castilla la Mancha | Centro de Investigación Agraria de Aguas Nuevas<br>Centro de Investigación Agraria de Albaladejito<br>Centro de Investigación, Experimentación y Servicios del Champiñón<br>Centro de Investigación Vitivinícola<br>Centro de Mejora Agraria "El Chaparrillo"<br>Centro Regional de Selección y Reproducción Animal<br>Centro de Experimentación de Almodóvar del Campo<br>Centro Regional Apícola |
| Cataluña | Instituto de Investigación y Tecnología Agroalimentarias (IRTA) |
| Comunidad Valenciana | Instituto Valenciano de Investigaciones Agrarias (IVIA) |
| Extremadura | Servicio de Investigación Agraria (SIA) |
| Galicia | Centro de Instituto Agrarias de Mabegondo (CIAM) |
| Madrid | Instituto Madrileño de Investigación Agraria (IMIA) |
| Murcia | Instituto Murciano de Investigación y Desarrollo Agrario y Alimentario (IMIDA) |
| Navarra | Instituto Técnico y de Gestión Agrícola (ITGA) |
| País Vasco | Instituto Tecnológico, pesquero y Alimentario (AZTI) |
| La Rioja | Centro de Investigación y Desarrollo Agrario (CIDA) |

Source: INIA.

| |
|---|
| Centro de Ciencias Medioambientales (CCMA) |
| Centro de Edafología y Biología Aplicada del Segura (CEBAS) |
| Estación Experimental Aula Dei (EEAD) |
| Estación Experimental Del Zaidin (EEZ) |
| Estación Experimental La Mayora (EELM) |
| Instituto de Agricultura Sostenible (IAS) |
| Instituto de Agrobiotecnología (IDAB) |
| Instituto de Ciencias Agrarias (ICA) |
| Instituto de Ganadería de Montaña (IGM) |
| Instituto de Investigaciones Agrobiológicas de Galicia (IIAG) |
| Instituto de Productos Naturales y Agrobiología (IPNA) |
| Instituto de Recursos Naturales y Agrobiología de Salamanca (IRNASA) |
| Instituto de Recursos Naturales y Agrobiología Sevilla (IRNAS) |
| Misión Biológica de Galicia (MBG) |
| Instituto de Agroquímica y Tecnología de Alimentos (IATA) |
| Instituto de Ciencia y Tecnología de Alimentos y Nutrición (ICTAN) |
| Instituto de Ciencias de la Vid y del Vino (ICVV) |
| Instituto de Fermentaciones Industriales (IFI) |
| Instituto de Investigación en Ciencias de Alimentación (CIAL) |
| Instituto de Investigaciones Marinas (IIM) |
| Instituto de la Grasa (IG) |
| Instituto de Productos Lácteos de Asturias (IPLA) |
| Instituto del Frío (IF) |

Source: CSIC.

**c) Technological Centers**

Technological centers constitute another type of entity, characterized by their involvement in activities to generate technological and R&D knowledge and develop applications. These centers are oriented toward the business world and productive outlets and dedicated to industrial research, experimental development, and innovation. They also foment an entrepreneurial culture and are meant to help companies in processes of internationalization.

| | |
|---|---|
| Castilla y León | Asociación de Investigación para la Mejora del Cultivo de la Remolacha Azucarera (AIMCRA) |
| | Fundación Centro Tecnológico Cereales Castilla y León (CETECE) |
| Comunidad Valenciana | Asociación de Investigación de la Industria Agroalimentaria (AINIA) |
| Extremadura | Asociación Empresarial de Investigación Centro Tecnológico Agroalimentario de Extremadura (CTAEX) |
| Galicia | Asociación de Investigación de la Industria Agroalimentaria (CECOPESCA) |
| Murcia | Asociación Empresarial de Investigación Centro Tecnológico |

|  | Nacional de la Conserva (CTC) |
|---|---|
| Navarra | Asociación Lechera de Vacuno y Ovino del País Vasco y Navarra (ALVO) |
|  | Centro Técnico Nacional de Conservas Vegetales (CTNVC) |

Source: Ministerio de Ciencia e Innovación.

## THE ROLE OF PROFESSIONAL ASSOCIATIONS

Professional associations have played an important role in Spain with regard to the process of information transfer to the agricultural and agro-alimentary sector. Along with the new democracy, a law was passed in 1977 (Law 19/1977) for the regulation of the right to freedom of syndication, permitting diverse associations dedicated to the defense of professional interests and those of their affiliates and producer organizations. The cooperative movement was also strengthened, resulting in a semi-public management of agrarian extension.

At present, the following associations are active:

- ASAJA (*Asociación Agraria de Jóvenes Agricultores*): founded in July 1989 with the fusion of three organizations —*Confederación Nacional de Agricultores y Ganaderos* (CNAG), *Unión de Federaciones Agrarias de España* (UFADE), and the *Centro Nacional de Jóvenes Agricultores* (CNJA). Some of these are considered offspring of the Agrarian Union and the Brotherhood of Farmers and Stockbreeders that date to the time of Franco. ASAJA represents its associates in the powerful Spanish Confederation of Business Organizations (*Confederación Española de Organizaciones Empresariales*, or CEOE) and integrates all agrarian activity: agriculture, livestock, forestry, environmental management and agro-tourism.

- COAG (*Coordinadora de Organizaciones de Agricultores y Ganaderos*): founded in 1977 and focusing largely on the realm of the agrarian cooperatives, which were considered natural economic structures in the rural setting. In 2008 a break-up occurred and the *Unión Confederal* was established, which took in several associations of a regional nature (*Unió de Pagesos de Cataluña*, *Ganaderos y Silvicultores de la Comunidad de*

*Madrid, Unión de Labradores y Ganaderos de la Comunidad Valenciana, Unión de Agricultores y Ganaderos de Extremadura, la Plataforma Agraria Libre de Canarias, la Unión de Agricultores,* and *la Unión de Campesinos de Castilla y León*).

- The Union of Small Farmers (*Unión de Pequeños Agricultores*, UPA): an organization rooted in the tradition of socialist agrarian syndicalism. The UPA represents the interests of proprietors of small or mid-sized farms and ranches and is integrated in the structure of the autonomous or self-employed workers of the General Workers' Union (*Unión General de Trabajadores*, UGT), one of the most important worker unions in Spain.

These associations defend the rights of their members before the Public Administrations and have a number of informational services at their disposal.

The cooperatives are businesses of the social economy. In Spain, co-ops of the agro-alimentary sector number some 4,000 and are represented in the Federation of Agro-alimentary Cooperatives.

In addition to professional associations and cooperatives, there is another type of organization with a more technical character. It is comprised of fruit and vegetable producers (created by Royal Decree, R.D. 1972/2008) and the inter-professional agro-alimentary associations that are regulated by Law 38/1994.

The inter-professional associations are organizations that represent businessmen in charge of the production, commercialization and transformation of a sector or a product of the agro-alimentary system, and their functions include promotion or research and the development and improvement of consumer information. In 2009 there were 25 such inter-professional associations in Spain, corresponding to the agricultural, livestock, forestry and fishing sectors. The inter-professional organizations may make obligatory agreements for their members and for all operators within the sector regarding matters such as product quality, environmental protection, market information and knowledge, public relations, and R&D&I.**AGRARIAN CHAMBERS**

We should also briefly describe the relevance of the Agrarian Chambers ~~as~~ to the associative movements appearing at the end of the 19th century. These originally arose to assist rural populations, with offices that would process diverse documentation. After long decades of uprising, in 1986 a law was passed (Ley 23/1986) to establish the legal basis of the Judicial Regime of the Agrarian Chambers, making them entities of the Public Right and giving them the means to act as consultative organs of the Public Administration, with authority to issue reports or studies. Yet they were not permitted to take on the functions of representing the professional and socioeconomic interests of farmers and ranchers, as these responsibilities still came under the professional organizations. Notwithstanding, due to the fact that in 2005 all the regions gained power regarding farming and livestock, and authority over the Agrarian Chambers, a law was passed (Law 18/2005) to repeal state control of the Chambers but not to abolish them; thus, the regions were given the power to maintain them or eliminate them.

The process of suppression of the Agrarian Chambers has been gradual since 2005, as they are now void of responsibility in the face of regional administrations that manage agricultural policies and alongside a conglomerate of business associations in the agrarian sector and cooperatives that represent the agricultural sector.

## COMMUNICATION OF AGRICULTURAL INFORMATION

A study of agricultural communication methods cannot overlook one of the earliest means of agricultural information dissemination: periodical publications, which might even be considered as the first modern form of agricultural information. In many cases periodicals recorded the prices of merchandise coming from ports and of the new crops or expressly aimed to educate farmers. The 18th century Enlightenment was the movement behind many of these publications in Spain, as before the appearance of the earliest periodical specializing in agriculture, the *Semanario de Agricultura y Artes* (1797-1808), there were numerous publications dealing with agricultural topics, such as *Diario Noticioso* or *Estafeta de Londres*. Their editors, normally illustrious writers, aspired to facilitate the nation's overall progress through their publication.

One of the most polemic periodicals was *El Censor* (1781-1787), which criticized the structures underlying land ownership. According to Acosta Meneses (2007, p. 645), *El Censor* became the most critical periodical of the era, as it "denounced the structure of land ownership as the main impediment in Spain for the progress of agriculture… Besides, it spoke of the misery of the farm laborers and of the high prices of land leases, as well as the opulence of the owners, who spent the benefits obtained on leisure, instead of re-investing them."

In 1797, Spain's very first publication strictly about agriculture appeared. The *Semanario de Agricultura y Artes dirigido a los párrocos* aspired to foment agrarian progress through the figure of the parish priest, who, in many populations, happened to be the only person able to read.

This time period also witnessed the appearance of the earliest publications about economic information. The *Semanario Económico* and the *Correo Mercantil de España y sus Indias* (1792-1808) were published by Gallard, an editor who declared that those persons dedicated to commerce should know what was going on around them and recommended they become somewhat familiar with "the storms, the harvests, the rise and fall of consumption, abundance and scarcity of fruit in the villages, the variety of their tastes, their different dealings and alliances, their economic providences, their mercantile enterprises, their progress in the arts" (Enciso Recio, 1958, p. 43-44). This can be seen as early evidence, at the end of the 18th century, of the activities that led to technological and economic awareness.

During this time, therefore, a separation of publications takes place; whereas some specialize in technical information and communications directed to farmers or producers, others focus on economic information, as their intended readership consists of those who commercialize the products (Acosta Meneses, 2007, p. 646).

Throughout the 19th century, information about the agricultural sector became generalized in the national press and many publications dedicated to the agrarian world were published by professional associations, such as those representing the agrarian and forestry engineers. Furthermore, *desamortización* ("de-amortization") occurred. This was a process by which uncultivated land was assessed —meaning that the number of proprietors and producers increased, thereby generating greater interest. The development of the stock market system and of other industrial sectors led to the appearance of financial information alongside the agricultural information.

The 20th century witnessed a change in how agricultural information was presented in the press depending on the political and social situation of the country. There were events and topics of great interest, such as the "Crisis of Subsistence" in 1905, the political projects deriving from the Agrarian Reform, the laws about rural or crop leasing, or the plans for colonization, along with other subjects and events of much lesser interest.

Agricultural information during the 20$^{th}$ century began to be associated with economic and political information, yet the 20$^{th}$ century also brought radio and television and, therefore, the diffusion of agricultural information over these new media. At present, various national, regional and local broadcasting networks sponsor agricultural programming on the radio. One such program is Agro Popular, which, since 1984, gathers and transmits information about the weather and market trends for agricultural or livestock products. On national public television (TVE) and on regional channels we find other examples. Noteworthy is the failed attempt to maintain a satellite channel specializing in agriculture: Agro Rural, directed to rural populations and professionals in the agricultural sector, stopped broadcasting in 2001.

All in all, one of the most important information sources in the mass media of Spain is an offshoot of the news agency EFE: EFEAGRO, an agency whose objective is to provide in-depth news and reports related to the agro-alimentary sector. They offer a general information service as well as services specializing in different sectors, such as fishing, wine production, and fruits and vegetables, among others. The news emitted by EFEAGRO may deal with national or community legislation; summaries of items from the regional, national or international press, EU news or norms, information related to the fishery or agro-fishery sectors, wholesale markets and Spain's exchange, international markets, ecology and the environment, agrarian economy, business, distribution, etc.**CONCLUSION**

The systems of agricultural information in Spain have been largely derived from initiatives of the Public Administration, with few private initiatives. Thus, the reality of Spain's Public Administration is necessarily reflected in the structure of these systems. The agricultural sector was extremely underdeveloped during the first half of the 20$^{th}$ century, making efforts undertaken in the following decades even more evident. The sector reflects the advantages and disadvantages of Spain's Autonomous System of community administration, a model in which regional authorities assume the main responsibilities in agricultural matters. In addition to the Public Administration, professional associations have played an important role in the dissemination of agricultural information in Spain.